\documentclass[11pt]{article}
\usepackage{graphicx}
\usepackage{amsmath}

\newcommand\pubnumber{DPF2015-247}
\newcommand\pubdate{\today}

\def\omegapinc{\textsuperscript{1}Omega-P, Inc, New Haven, CT 06511, USA; 
\textsuperscript{2}Yale University, New Haven, CT, USA; 
\textsuperscript{3}Fermi National Accelerator Laboratory, Batavia, IL 60510;\\
\textsuperscript{4}Calabazas Creek Research, Inc., San Mateo, CA, 94404;\\ 
\textsuperscript{5}Communication \& Power Industries, LLC., Palo Alto, CA, 94304}
\def\support{\footnote{This work was supported by several SBIR grants DE-SC0000926, DE-SC0000927, DE-SC0007532 and 
DE-SC0007533  to Omega-P, Inc. from Department of Energy, Office of High Energy Physics }}

\def\Title#1{\begin{center} {\Large #1 } \end{center}}
\def\Author#1{\begin{center}{ \sc #1} \end{center}}
\def\Address#1{\begin{center}{ \it #1} \end{center}}

\newcommand\pubblock{\rightline{\begin{tabular}{l} \pubnumber\\
         \pubdate  \end{tabular}}}
\newenvironment{Abstract}{\begin{quotation}  }{\end{quotation}}
\newenvironment{Presented}{\begin{quotation} \begin{center} 
             PRESENTED AT\end{center}\bigskip 
      \begin{center}\begin{large}}{\end{large}\end{center} \end{quotation}}

\def\beq{\begin{equation}}
\def\eeq#1{\label{#1}\end{equation}}
\def\eeqn{\end{equation}}

\def\beqa{\begin{eqnarray}}
\def\eeqa#1{\label{#1}\end{eqnarray}}
\def\eeqan{\end{eqnarray}}

\let\bar=\overbar

\def\Dslash{\not{\hbox{\kern-4pt $D$}}}
\def\dslash{\not{\hbox{\kern-2pt $\del$}}}

\def\msb{{\bar{\ssstyle M \kern -1pt S}}}

\begin{document}
\begin{titlepage}
\pubblock

\vfill
\Title{Compact Low-Voltage, High-Power, Multi-beam Klystron for ILC: Initial Test Results\support
}
\vfill
\Author{ V.E. Teryaev\textsuperscript{1}, S.V. Shchelkunov\textsuperscript{1,2} (sergey.shchelkunov@gmail.com),\\
S.Yu. Kazakov\textsuperscript{3}, J.L. Hirshfield\textsuperscript{1,2}, \\
R.L. Ives\textsuperscript{4}, D. Marsden\textsuperscript{4}, G. Collins\textsuperscript{4},\\
R. Karimov\textsuperscript{4}, R. Jensen\textsuperscript{5}}
\Address{\omegapinc}
\vfill
\begin{Abstract}
Initial test results of an L-band multi-beam klystron with parameters relevant for ILC are presented.  The chief distinction of this tube from MBKs already developed for ILC is its low operating voltage of 60 kV, a virtue that implies considerable technological simplifications in the accelerator complex. To demonstrate the concept underlying the tube’s design, a six-beamlet quadrant (a 54 inch high one-quarter portion of the full 1.3 GHz tube) was built and recently underwent initial tests, with main goals of demonstrating rated gun perveance, rated gain, and at least one-quarter of the full 10-MW rated power. Our initial three-day conditioning campaign without RF drive (140 microsec pulses @60 Hz) was stopped at 53\% of full rated duty because of time-limits at the test-site; no signs appeared that would seem to prevent achieving full duty operation (i.e., 1.6 msec pulses @10 Hz). The subsequent tests with 10-15 microsec RF pulses confirmed the rated gain, produced output powers of up to 2.86 MW at 60 kV with high efficiency and 56 dB gain, and showed acceptable beam interception.  These results suggest that a full version of the tube should be able to produce up to 11.5 MW.  Follow-on tests are planned for later in 2015.
\end{Abstract}
\vfill
\begin{Presented}
DPF 2015\\
The Meeting of the American Physical Society\\
Division of Particles and Fields\\
Ann Arbor, Michigan, August 4--8, 2015\\
\end{Presented}
\vfill
\end{titlepage}
\def\thefootnote{\fnsymbol{footnote}}
\setcounter{footnote}{0}

\section{Introduction}

Mulit-beam klystrons (MBKs) are the next stage of klystron evolution~\cite{ref1},~\cite{ref2}. The distinct feature of MBKs is their relatively low-voltage beams (beam-lets) as compared to the beam in the traditional tubes. It delivers numerous advantages: MBKs can use a simpler (and thus, a cheaper) modulator, no pulse-transformer is required; and there is no need for a high-voltage oil tank surrounding the gun. The MBK tube can be made significantly shorter (e.g. by a factor at least 2) than a single-beam version delivering the same power and pulse; in particular the collector length can be greatly reduced. The challenges associated with MBKs are typically to design the cavity chain where all parasitic modes are sufficiently detuned, which is accomplished by using a set of metal features inserted into the cavity regions to exercise either inductive or capacitive detuning such as in the input and/or output cavities where the RF field interacts with all the beam-lets (a set allows to perform detuning in a manner so that conditions when one beam-let interacts somewhat differently with the operating mode than another beam-let are not created), or by using individual gain and penultimate cavities each serving a sub-group of beam-lets (beam-let cluster) and working at the fundamental mode. To easily match the beam between the gun, the cavities, and collector, the magnetic system is typically divided by iron pole-pieces into regions of independent control. An associated challenge is to have coil and pole-piece configuration where the transverse magnetic field on the axis of each beam-let is less than 0.5\% of the longitudinal field, while keeping the magnetic system design relatively simple. In ~\cite{ref1} and references therein, detailed descriptions can be found of tools, method and approaches to address the aforementioned challenges; in particular, the contemporary state-of-art codes such as MAGIC~\cite{ref4}, MERMAID~\cite{ref5}, and DGUN~\cite{ref6} are invaluable in designing MBKs.

\section{Multi-beam klystron performances}
 
\begin{figure}
 \centering
 \begin{minipage}{4.75in}
 \centering
 \includegraphics[width=4.75in]{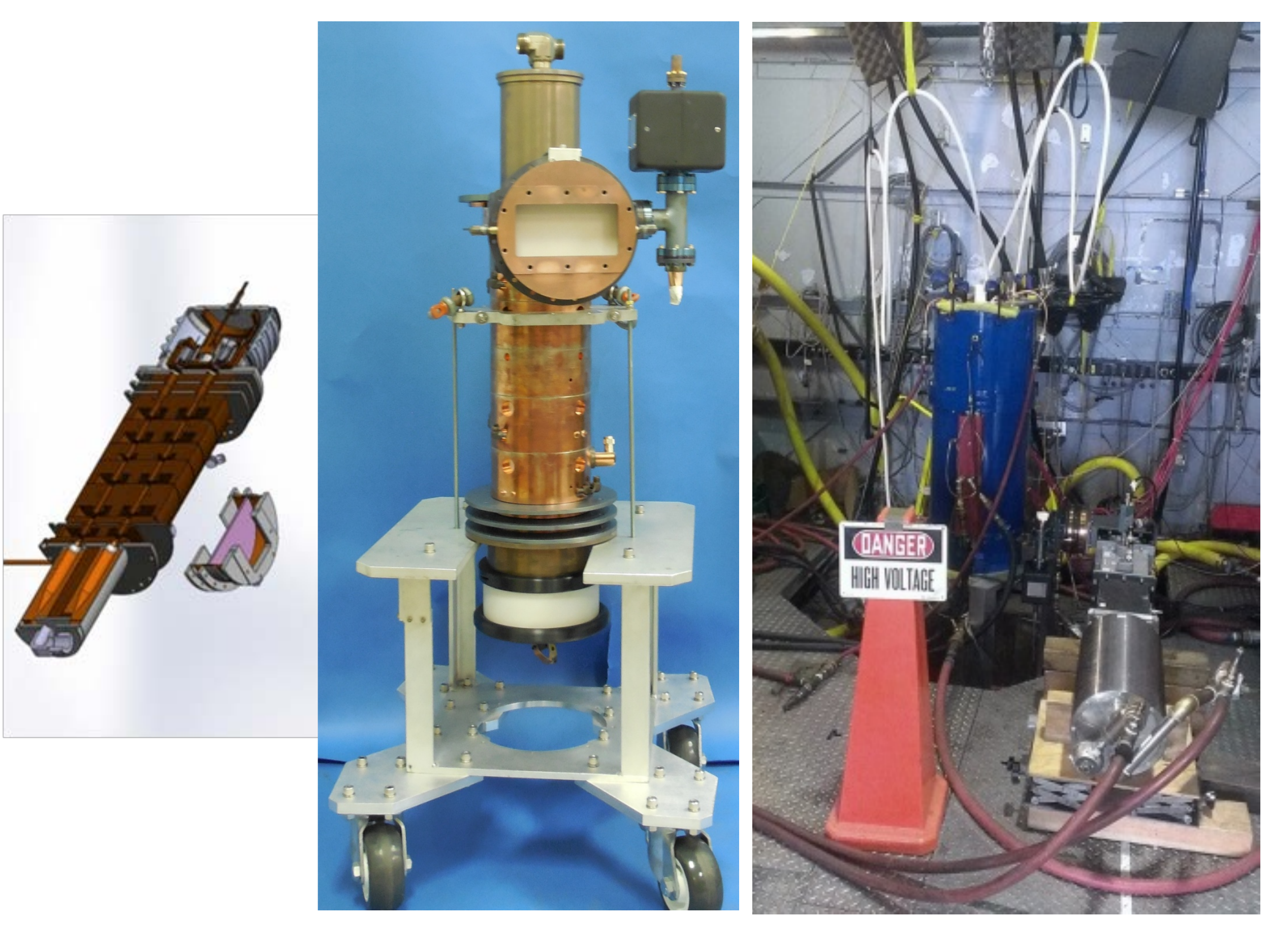}
 \end{minipage}
 \caption{\textbf{(left)} a 3D cut-view of the RF structure; \textbf{(middle)} photo showing the RF structure before it was inserted the coil-assembly.  \textbf{(Right)} the test-setup – the klystron, RF-window and RF-load - as assembled at the CPI test-bench (with the gun facing up).}
 \label{fig:1}
 \end{figure}

 Figure~\ref{fig:1} shows some cut-views and photos.
 
\begin{figure}
 \centering
 \begin{minipage}{2.9in}
 \centering
 \includegraphics[width=2.9in]{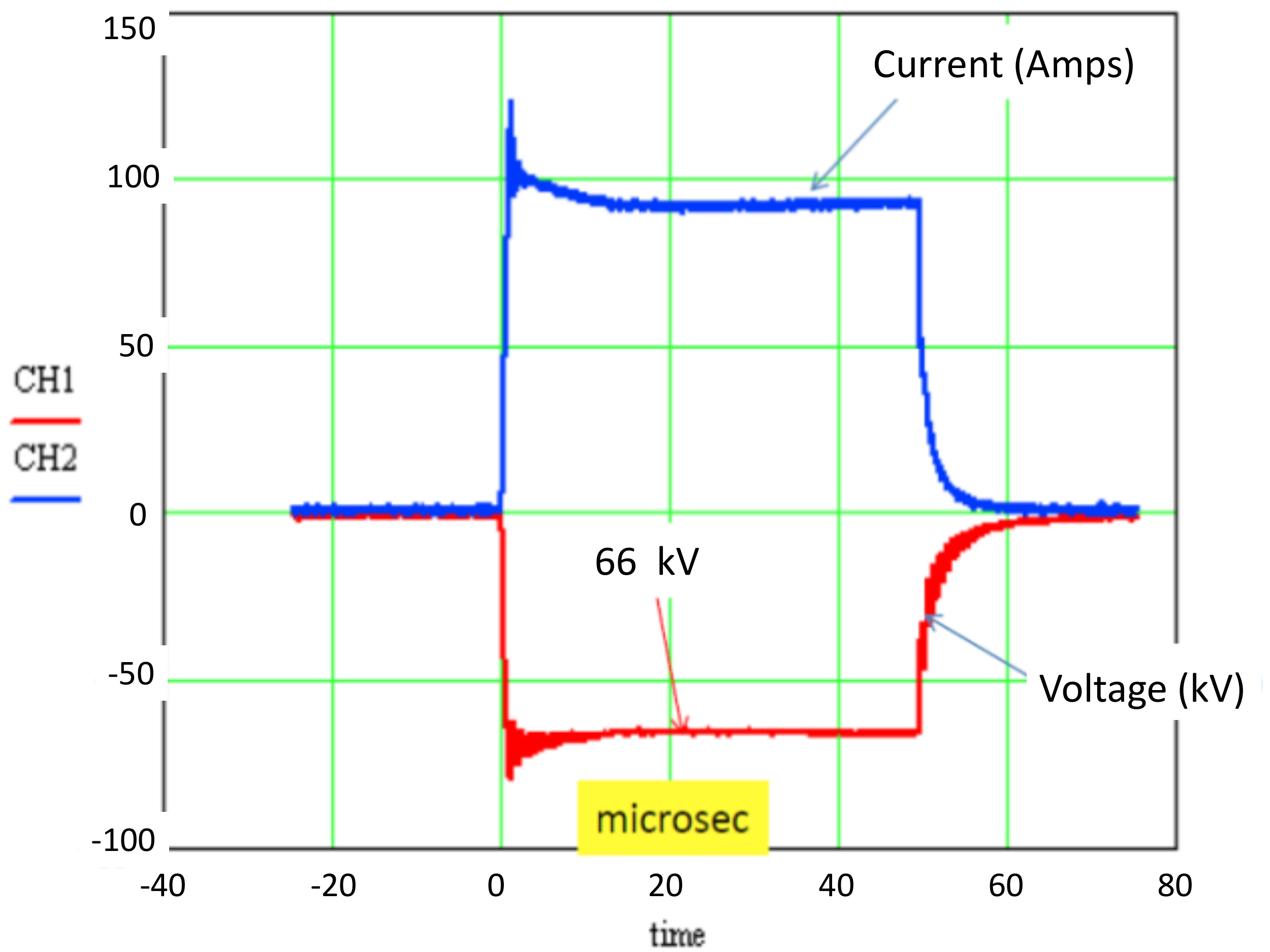}
 \end{minipage}
 \begin{minipage}{2.9in}
 \centering
 \includegraphics[width=2.9in]{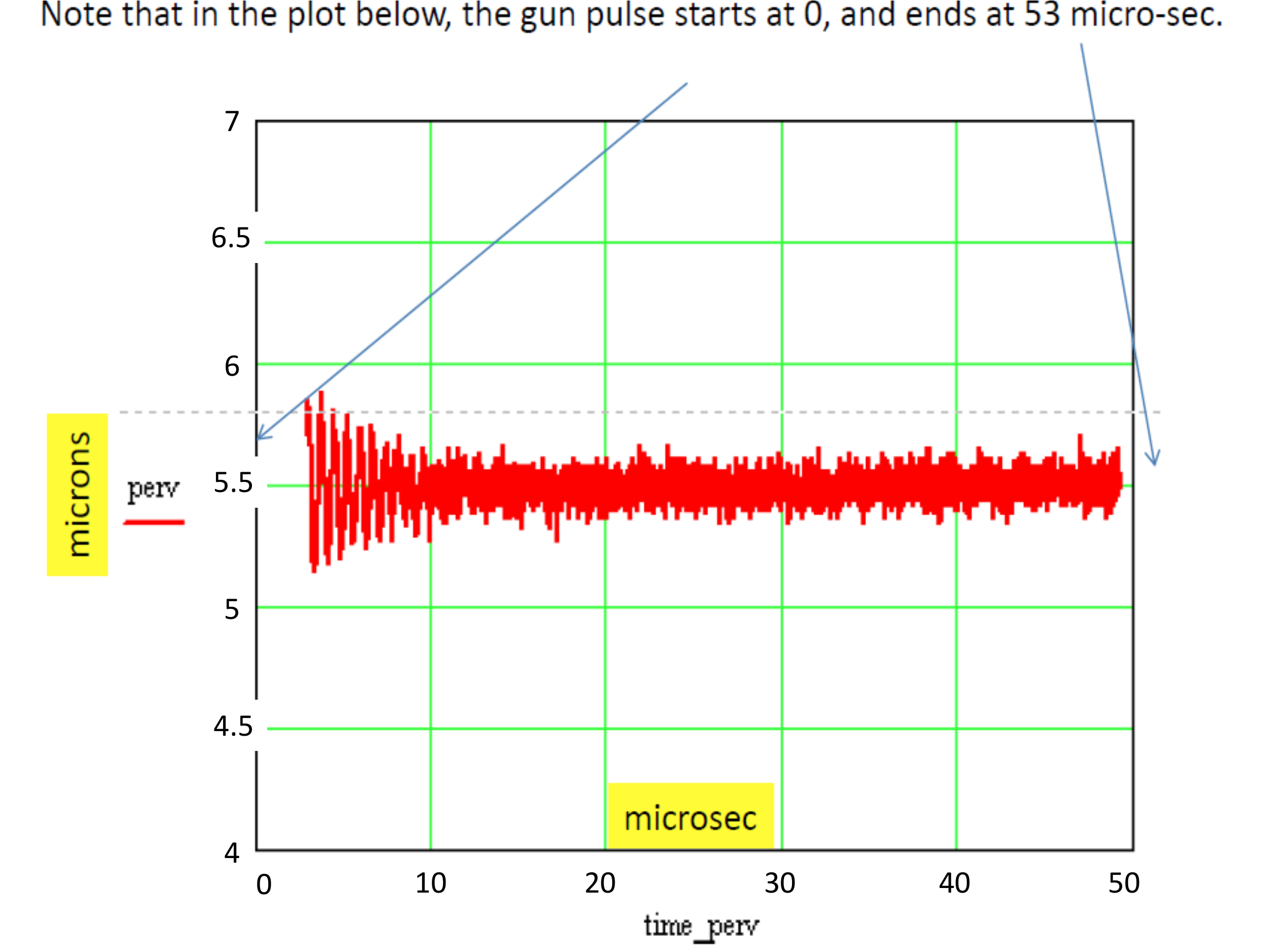}
 \end{minipage}
 \caption{\textbf{(left)} gun current (50 A/div) and voltage (50 kV/div) vs. time (20 $\mu$sec/div),  \textbf{(right)} gun micro-perveance (5.5$\mu$A/V\textsuperscript{3/2}) vs. time(10 $\mu$sec/div) at 66kV, as recorded between 0 and 53 $\mu$sec, which was the duration of the pulse-top in this particular case.}
 \label{fig:2}
 \end{figure}

 We start with the description of the gun performances. The gun produces 6 beams equidistantly spaced around the common bolt circle. The measured perveance (from 35-60kV) was found to be 5.5$\cdot$ 10\textsuperscript{-6} A/V\textsuperscript{3/2}, which is somewhat higher than the design value 4.9$\cdot$ 10\textsuperscript{-6} A/V\textsuperscript{3/2}. The difference between the designed value and the measured value is yet to be explained. The emission test results showed that the required filament current is 53.8-54 Amps at 10 VAC at 60Hz. Figure~\ref{fig:2} shows the behavior of the 6 beam-let gun voltage and full current vs. time (left), and also the perveance vs. time recorded at 66kV (the shown plot is a typical behavior of the perveance at different gun voltages).
 
 Our initial three-day conditioning campaign without RF drive (140 microsec pulses @60 Hz) was stopped at 53\% of full rated duty because of time-limits at the test-site; no signs appeared that would seem to prevent achieving full duty operation (i.e., 1.6 msec pulses @10 Hz).
 
 \begin{figure}
 \centering
 \begin{minipage}{4.25in}
 \centering
 \includegraphics[width=4.25in]{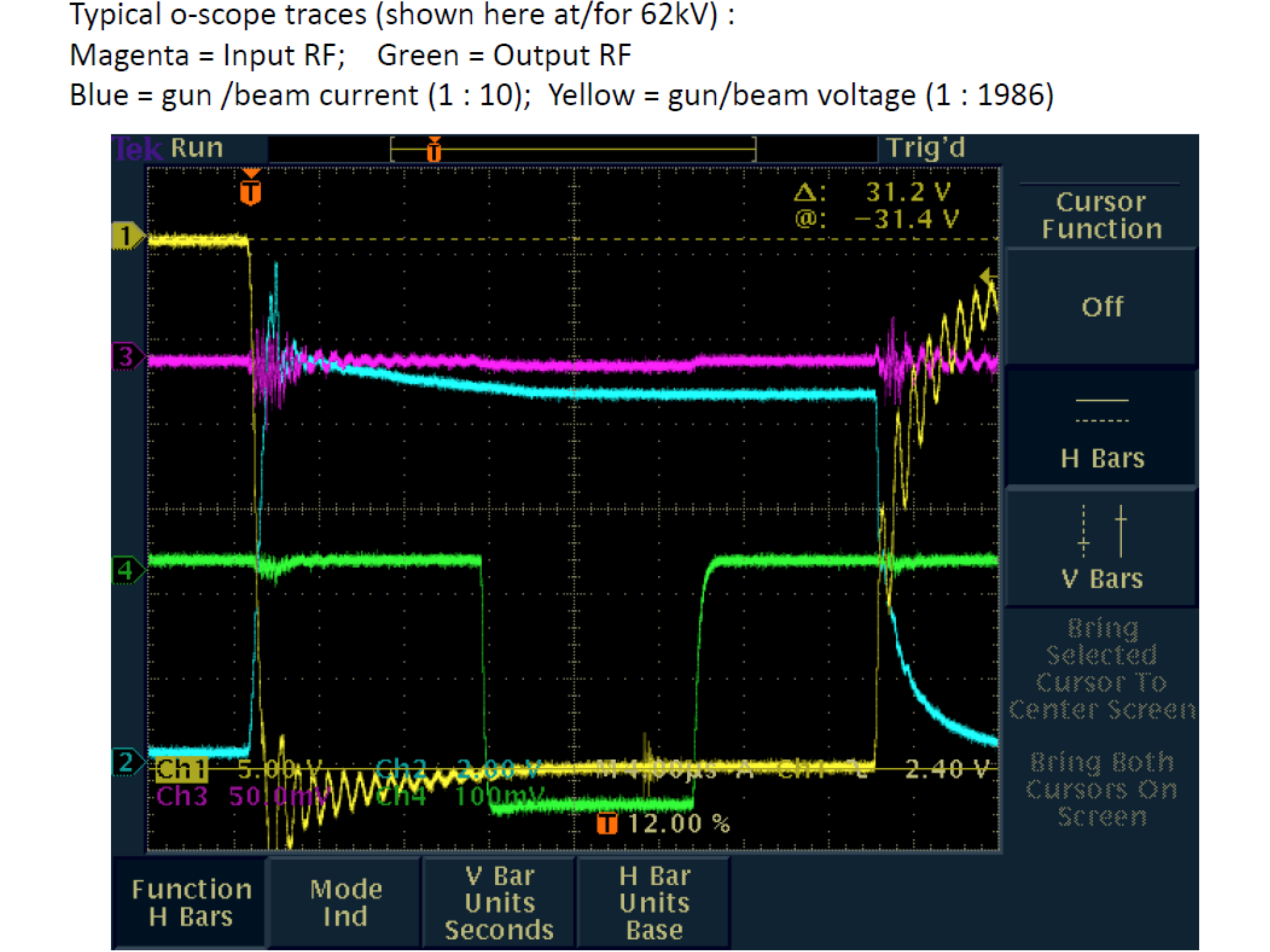}
 \end{minipage}
 \caption{typical pulses (the horizontal scale is 5 $\mu$sec/div). Here, the shown RF-pulse length is $\sim$12.5 $\mu$sec.}
 \label{fig:3}
 \end{figure}
 
 \begin{figure}
 \centering
 \begin{minipage}{2.9in}
 \centering
 \includegraphics[width=2.9in]{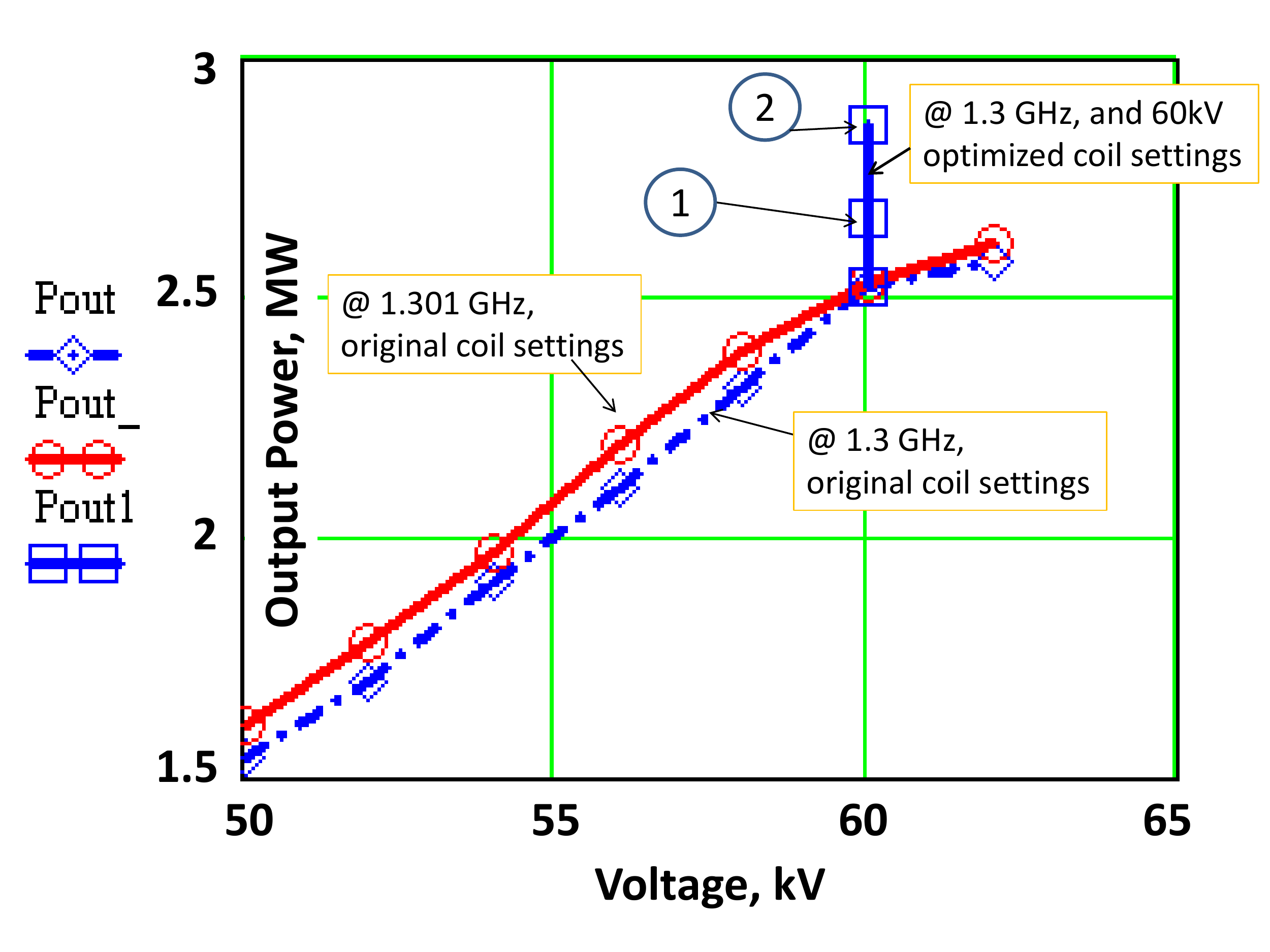}
 \end{minipage}
 \begin{minipage}{2.9in}
 \centering
 \includegraphics[width=2.9in]{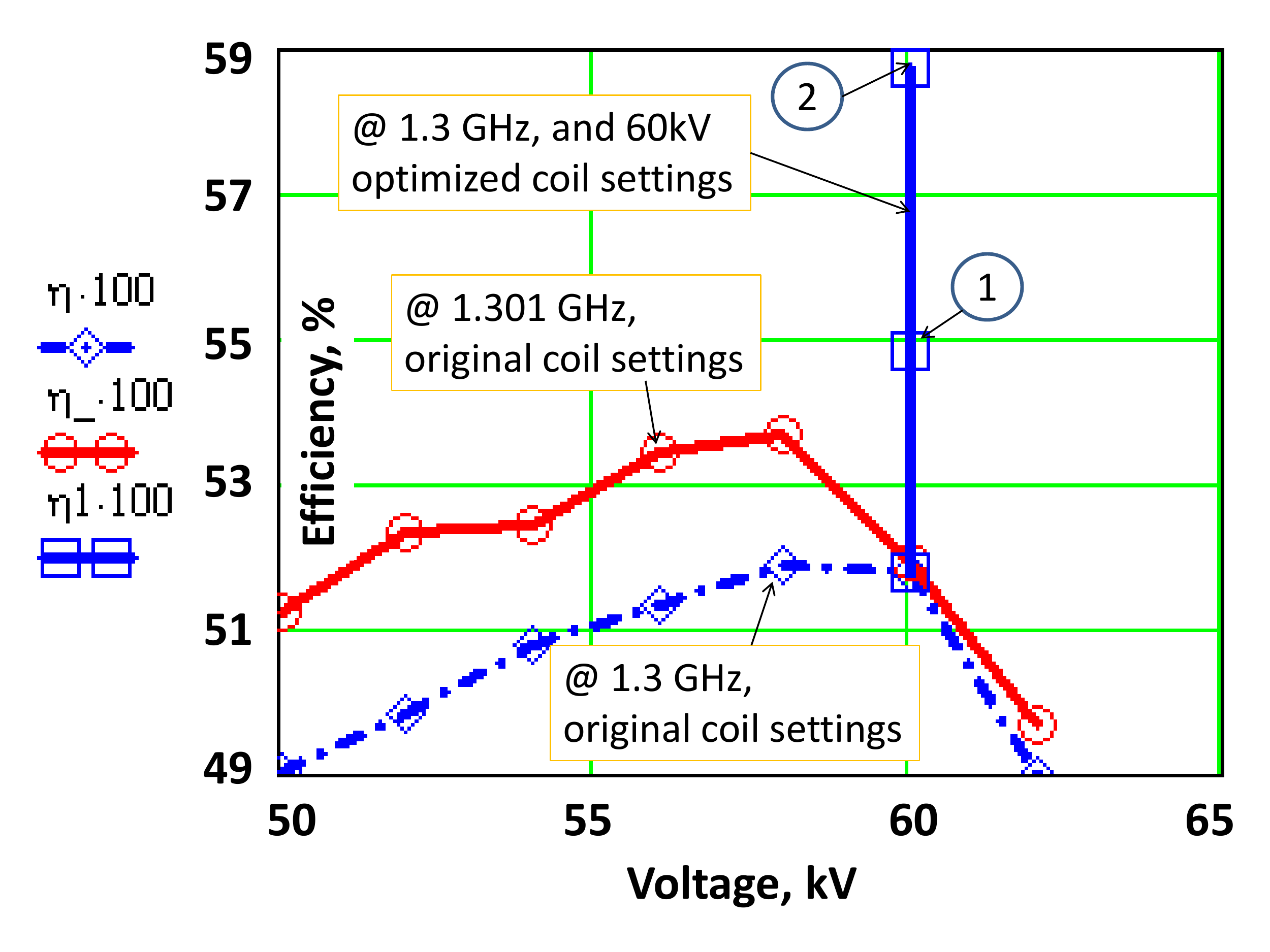}
 \end{minipage}
 \caption{\textbf{(left)} the output power (0.5 MW/div,\textit{via calorimetry on a matched load}) versus the beam-voltage (5 kV/div). \textbf{(Right)} the efficiency (\% of the total beam-power)}
 \label{fig:4}
 \end{figure}
 
 \begin{figure}
 \centering
 \begin{minipage}{3.25in}
 \centering
 \includegraphics[width=3.25in]{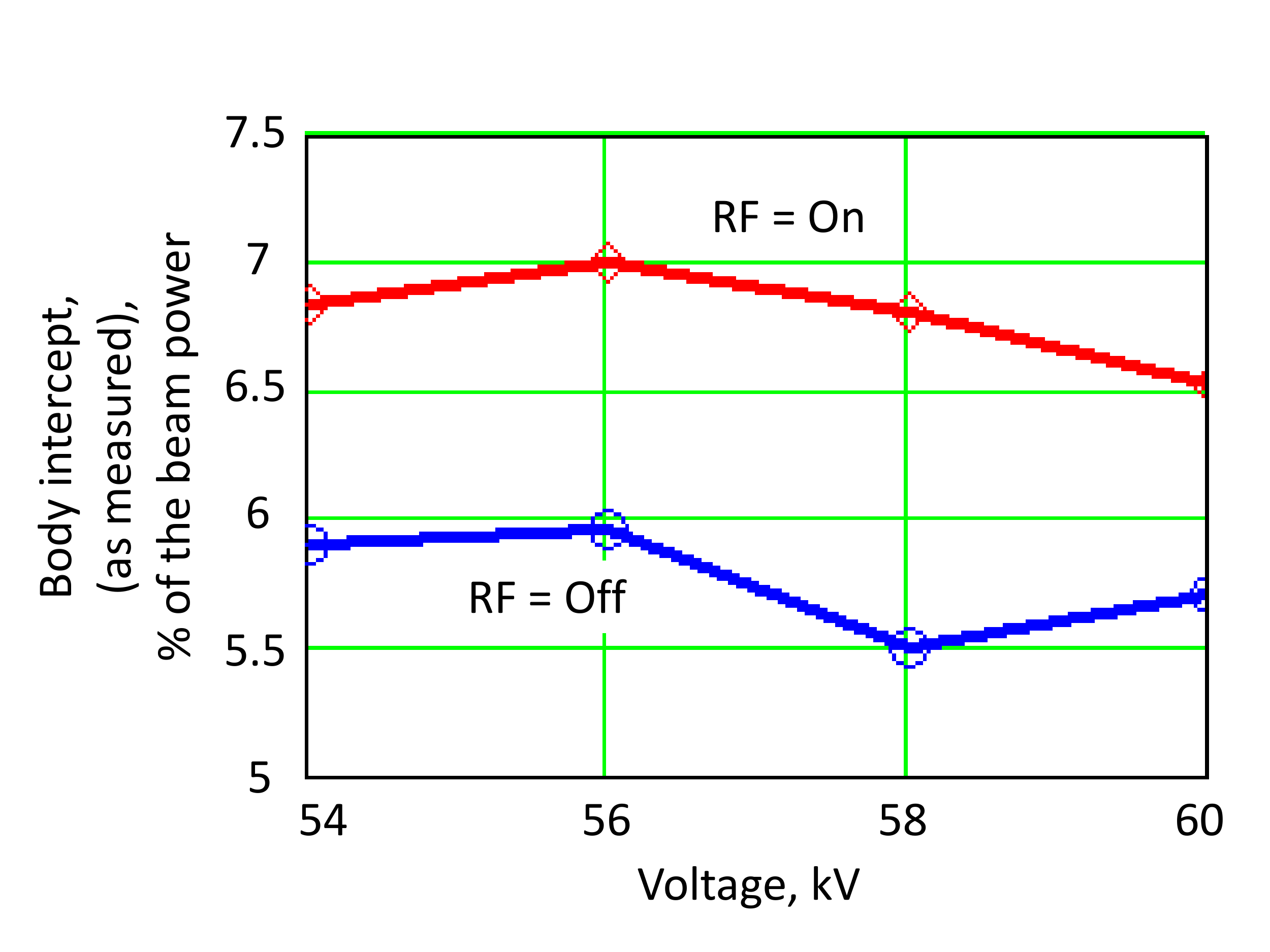}
 \end{minipage}
 \caption{ The body intercept via calorimetric measurements on the klystron body (coil settings are as per “Optimized 1” set). The intercept is shown in the percentage of the beam power.}
 \label{fig:5}
 \end{figure}
 
 \begin{figure}[h]
 \centering
 \begin{minipage}{2.9in}
 \centering
 \includegraphics[width=2.9in]{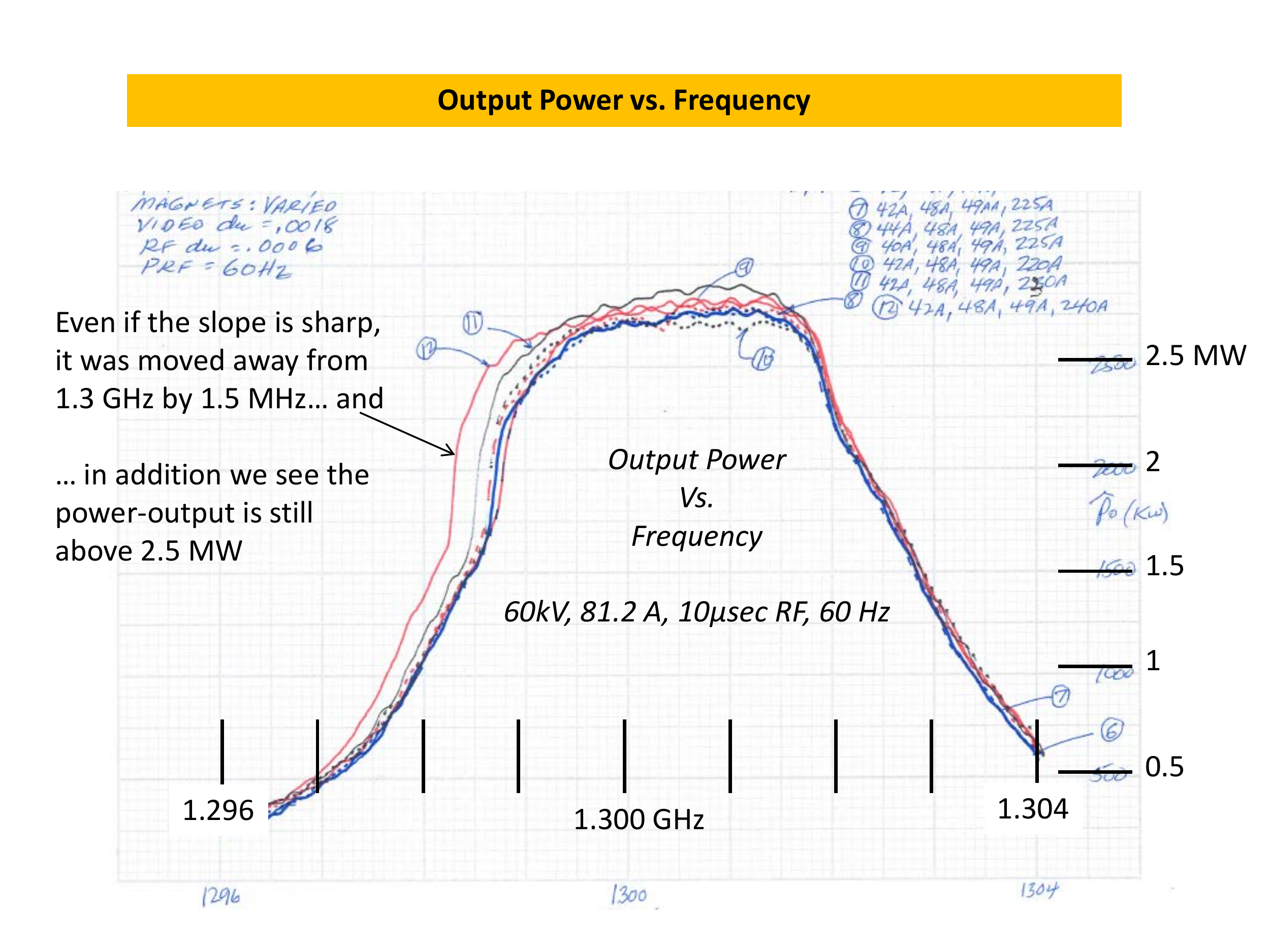}
 \end{minipage}
 \begin{minipage}{2.9in}
 \centering
 \includegraphics[width=2.9in]{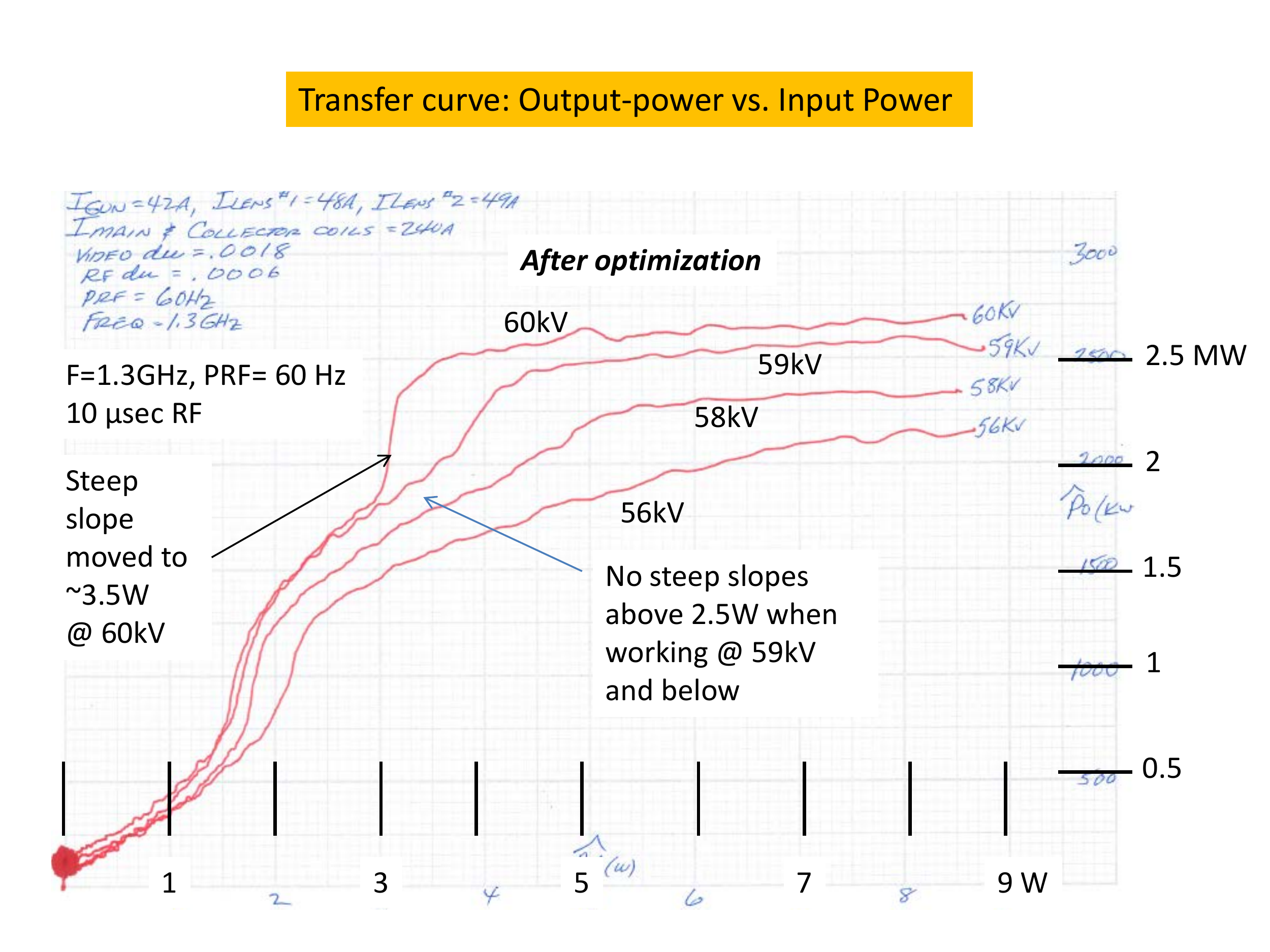}
 \end{minipage}
 \caption{\textbf{(left)} example of investigations of the bell-jar shape of the dependence of the Output-power vs. the RF frequency at 60kV vs. different coil settings. \textbf{(Right)} An example of a family of transfer curves for the coil settings that minimized the body-interception. [~Input power is in Watts (horizontal axis) ]. The coils are set to “Optimized 1”}
 \label{fig:6}
 \end{figure}
 The RF-structure has 6 beam-tunnels. The structure has 6 cavities: the input, gain, second-harmonic, penultimate-\#1, penultimate-\#2, and the output. There is one output port, and one RF-window employed to get the power out. The output flange is WR-650.
 The RF tests were carried out with up to 15 microsec long pulses. Figure~\ref{fig:4} shows the output power and efficiency measured for different coil settings. There are 4 coils to guide and match the beam, named the gun-coil, lens-\#1, lens-\#2, and the main-coil. The gun coil surrounds the gun region. Lens-\#1 and -\#2 are to adjust the magnetic field in the region between the gun and the RF-structure. The main-coil controls the magnetic field along the RF-structure. The settings are listed in the Table~\ref{tab:1}. The produced power (@60kV) was recorded to be as shown in Table~\ref{tab:2}. In this table, the cited body intercept is simply a calorimetric measurement on the klystron body. It requires further post-processing to find the actual amount of the beam current intercepted by the body for there is a heat transfer between the different klystron components to be taken into account.

\begin{table}
\caption{coil settings (see performances in Fig~\ref{fig:4} (above) and Table~\ref{tab:2} (below)}
\begin{tabular}{c c c c c}
\hline
Name of the set & Gun-coil (Amps) & Lens \#1 (Amps) & Lens \#2 (Amps) & Main coil (Amps)\\ \hline
Original & 42 & 45 & 45 & 225 \\
Optimized 1 & 42 & 48 & 49 & 240 \\
Optimized 2 & 42 & 48 & 49 & 230 \\ \hline
\end{tabular}
\label{tab:1}
\end{table}

\begin{table}
\caption{measured output power (\textit{calorimetry on a matched load}), and body intercept (\textit{calorimetry on the klystron body}) @ 60kV} 
\begin{tabular}{c c c c}
\hline
Name of the set & \textbf{Original} & \textbf{Optimized 1} & \textbf{Optimized 2} \\ \hline
\textbf{Output Power, MW} & 2.53 & 2.63 - 2.67 & 2.86 \\
\textbf{Efficiency, \% of the beam-power} & 52 & 54-55 & 59 \\
\textbf{Body intercept, \% of the beam-power} & 10 & 6.5 & not recorded \\ \hline
\end{tabular}
\label{tab:2}
\end{table}

The body intercept measured via calorimetry on the klystron body is shown in Fig~\ref{fig:5}. These measurements were done having the coil settings as per “Optimized 1” set. The same consideration as for the Table~\ref{tab:2} applies here.

Studies were conducted to investigate and the bell-jar shape of the dependence of the Output-power vs. the RF frequency to have the smooth slopes, and similarly for the dependence of the Output-power vs. the Input/Drive-power (transfer curve). Figure~\ref{fig:6} (left) demonstrates a family of curves showing the dependences of the Output-power vs the RF frequency for different coil settings. This exercise led, in particular, to obtaining the “Optimized 1” coil set (mentioned above). Figure~\ref{fig:6} (right) shows an example of a family of transfer curves for the coil settings (Optimized 1) that minimized or nearly minimized the body-interception, and improved the bell-jar shape presented before by the red curve on the left hand-side. One observes that, in particular, @ 59kV, a reasonably smooth transfer curve results starting at 2W -  2.5W of the drive power. At the same time, 2.5 MW or even more MWs are achievable once the drive power $\geq$ 5 W, bringing the performance (power-wise) above the design specs.

\section{Conclusions}

The klystron demonstrated already that with 10-15 microsec RF pulses, output power of up to 2.86 MW at 60 kV with high efficiency and 56 dB gain can be produced. The test were performed at the repetition rate as high as 60Hz. The achieved maximum efficiency was 59 \%. Given the numbers, it can be speculated that if further optimizations of the coil settings had led to nearly zero beam intercept, the efficiency could have been as high as 65 \%.     Lastly, given all the test results, no signs appeared that would seem to prevent achieving full duty operation (i.e., 1.6 msec pulses @10 Hz).


\begin{thebibliography}{99}

\bibitem{ref1}
"Low Beam Voltage, 10MW, L-Band Cluster Klystron", V.E. Teryaev, et al., Procs of PAC09, Vancouver, BC, Canada, and refs. therein;
\bibitem{ref2}
"Status of High-Power Low-Voltage Multi-Beam Klystrons for ILC and Project X", by V.E. Teryaev (Omega-P, Inc, New Haven, CT), S.V. Shchelkunov (Yale University, New Haven, CT), J.L. Hirshfield (Omega-P, Inc, New Haven, CT and Yale University, New Haven, CT), N. Solyak (Fermilab, Batavia, IL), V. Yakovlev (Fermilab, Batavia, IL) S. Yu. Kazakov (Fermilab, Batavia, IL), in AAC 2014 Procs, ( The 16th Advanced Accelerator Concepts Workshop (AAC 2014),  San Jose, CA, July 13 - 18, 2014 )
\bibitem{ref3}
Omega-P, Inc. FINAL REPORT "RF CAVITY CHAIN AND MAGNETIC CIRCUIT FOR A 10-MW, 1.3-GHz, LOW-VOLTAGE MULTI-BEAM KLYSTRON", Based on efforts under Phase II grant DE-SC0000927, prepared by J. L. Hirshfield, Principal Investigator, tel:  (203) 789-1164,   e-mail:  jay@omega-p.com
\bibitem{ref4}
MAGIC, User’s Manual, see e.g. Mission Research Corp. MRC/WDC-R-409, 1997;
\bibitem{ref5}
MERMAID, User’s Guide see e.g. SIM Limited, 2005
\bibitem{ref6} 
"DGUN-code for simulation of intense axial-symmetric electron beams", p. 172, 6th Inter. Comp. Accel. Physics Conf, TU Darmstadt, Germany, 2000

\end{thebibliography}
\end{document}